\documentclass{ws-procs975x65}
\usepackage{graphicx} 
\usepackage{amsfonts, amsmath, amssymb}
\def\beq{\begin{equation}}
\def\eeq{\end{equation}}

\begin{document}

\title{Developing Tools for Multimessenger Gravitational Wave Astronomy}
\author{Maria C. Babiuc Hamilton$^*$}

\address{Department of Physics, Marshall University,\\
Huntington, WV 25755, US\\
$^*$E-mail: babiuc@marshall.edu}

\begin{abstract}
The Marcel Grossmann triennial meetings are focused on reviewing developments in gravitation and general relativity, aimed at understanding and testing Einstein's theory of gravitation. 
The $15^{th}$ meeting (Rome, 2018) celebrated the $50^{th}$ anniversary of the first neutron star discovery (1967), and the birth of relativistic astrophysics. 
Another discovery of the same caliber is the detection of the binary neutron star GW170817 in 2017 -- almost as if to celebrate the same jubilee -- marking the beginning of multi-messenger gravitational wave astronomy. 
We present work in progress to craft open-sourced numerical tools that will enable the calculation of electromagnetic counterparts to gravitational waveforms: the {\tt GiRaFFE} (General Relativistic Force-Free Electrodynamics) code.
{\tt GiRaFFE} numerically solves the general relativistic magnetohydrodynamics system of equations in the force-free limit, to model the magnetospheres surrounding compact binaries, in order (1) to characterize the nonlinear interaction between the source and its surrounding magnetosphere, and (2) to evaluate the electromagnetic counterparts of gravitational waves, including the production of collimated jets.
We apply this code to various configurations of spinning black holes immersed in an external magnetic field, in order both to test our implementation and to explore the effects of (1) strong gravitational field, (2) high spins, and (3) tilt between the magnetic field lines and black hole spin, all on the amplification and collimation of Poynting jets. 
 We will extend our work to collisions of black holes immersed in external magnetic field, which are prime candidates for coincident detection in both gravitational and electromagnetic spectra.

\end{abstract}


\bodymatter


\section{Introduction}
The 2017 detection of both gravitational waves (Ref.~\refcite{GW170817}) and electromagnetic radiation (Ref.~\refcite{GW170817-EM}) from two colliding neutron stars is a very important event, akin to the discovery of neutron stars in 1967 by J. Bell and A. Hewish.  
This event -- the rightfully-named Golden Binary -- started the gold rush of multi-messenger gravitational wave astronomy. 
Observations from neutron star mergers provide deep insights into the highly-nonlinear interaction between strong gravity and the surrounding magnetosphere, which could explain the mechanism behind the emission of short gamma ray bursts and relativistic jets.
Sources of strongly collimated astrophysical jets come at all scales, from pulsars to quasars, but they have in common one fact: they are compact objects rotating within their magnetospheres. 
Spinning neutron stars have a dipolar magnetic field and emit electromagnetic radiation due to the tilt between the spin axis and the magnetic field lines.
Black holes don't have magnetic poles, and therefore their magnetic fields are external, being generated by the accretion disc. 
Although black holes tend to align their spin with the angular momentum of the accretion disc, the spin and orientation of the accretion disks usually change as the holes grow through mergers and gas accretion. 
\cite{1003.4404}
Initially, merging supermassive black holes don't rotate in the same direction, and misalignment between the accretion disc and the black hole spin is expected. 

Recent magnetohydrodynamic simulations show that for short timescales, jets are only partially aligned with the black hole spin.\cite{1707.06619}
The influence of the tilt between the black hole spin and the rotation axis of the accretion disc, and therefore the role the orientation of the magnetic field plays in the formation and quenching of jets, is not yet fully understood.\cite{1506.04056}
Another process that is not completely understood is what is under the hood of the main engine that powers the jet. 
Are high black hole spins essential in triggering jet production, or is the most important role played by the dynamics of the accretion disc?\cite{1201.4163}
Theoretical models show that relativistic jets can be generated from the rotational energy of a rapidly spinning black hole in strong magnetic fields through the Blandford-Znajek (BZ) mechanism.\cite{1307.4752}
The vacuum around a rotating black hole is electromagnetically active, and gives rise to strong electric fields when the black hole is surrounded by a magnetic field.
Those gravito-rotationally induced electric fields cause a toroidal magnetic field, which plays an important role in collimating astrophysical jets.
Astronomical observations report that the transition from radio-quiet to radio-loud quasars occurs for black hole spins around $0.7$ to $0.9$ times the speed of light.\cite{1710.00316}
If black hole spin is at the root of this transition, one should see a sharp delimitation in the value of the black hole spins necessary for the onset and quenching of the jet.\cite{1710.01440}
Another problem arises when two merging black holes, embedded in an external magnetic field, spiral towards each other. 
The gravitational waves produced in this astrophysical scenario are expected to exert a direct effect on the magnetic fields. This coupling between the magnetic field and the surrounding dynamical spacetime induces electromagnetic waves, coincidental with the production of gravitational waves, which provides a detection avenue for multimessenger gravitational wave astronomy.\cite{1602.01492}

We engage in numerical explorations to study the interplay between black holes and the surrounding magnetosphere, in order to understand how spinning black holes drag the space-time around them, causing the collimation and amplification of jets.
This is a problem not yet completely elucidated, although great strides have been made towards it's resolution.\cite{komissarov, 0911.3889}
 We use the recently-released open-source {\tt GiRaFFE} code (Ref.~\refcite{GIR}) to perform numerical simulation of single Kerr black holes evolving in external magnetosphere and to look at the effect of black hole spin on the output of electromagnetic luminosity in order to discern if there is a threshold spin for the production or extinction of collimated jets.
Next we study the effect of tilted magnetospheres around a black hole with a fixed spin of $0.8$ times the speed of light, in order to discern if the tilt angle plays any role in the amplification or reduction of the jet power.

 
\section{Methods} 
\emph {Theoretical Approach}:
In order to explore the coupling between highly energetic gravitational and electromagnetic fields
we use the General Relativistic Force-Free Electrodynamics ({\tt GRFFE} ) ansatz.
{\tt GRFFE} is a limiting case of General Relativistic MagnetoHydroDynamics (GRMHD), consisting on ideal plasma coupled to strong electromagnetic and gravitational fields, when the magnetic field energy dominates and the fluid pressure can be ignored.
This theoretical model combines the assumption of perfect electric conductivity of the plasma with the fact that the Lorentz force in the comoving frame is immediately neutralized by induced electric currents, and it is commonly implemented in numerical codes to analyze the magnetized environment around pulsars or accreting black holes.~\cite{1310.3274}
This {\tt GRFFE} approach is implemented in our open-source {\tt GiRaFFE} code (Ref.~\refcite{etienne17}) designed to model the magnetosphere of highly-relativistic objects, such as neutron stars and black holes.
The spacetime is evolved with the Einstein field equations, while the dynamics of the magnetized fluid are modeled using the conservation law $\nabla_{\nu}T_{EM}^{\nu \mu}=0$ for the energy-momentum tensor $T_{EM}^{\nu \mu}$, and the Maxwell's equations $\nabla_{\nu}^*F^{\nu \mu}=0$ for the electromagnetic tensor $F^{\nu \mu}$ in the force-free limit, $F^{\nu \mu}J_{\nu}=0$. 
The updated variables are the magnetic field $B^i$ and the Poynting vector 
$S_i=-\gamma_{i \nu}n_{\mu}T_{EM}^{\nu \mu}$.\cite{etienne17} 
Here, $J_{\nu}$ is the 4-current, $\gamma_{i \nu}$ is spatial three-metric, and $n_{\mu}$ the timeline unit vector.

\emph {Numerical Techniques}:
The code evolves the vector potential $A_\mu = (\Phi, A_i)$ and Poynting vector $S_i$ one-forms, supplemented with an electromagnetic gauge evolution equation on a staggered grid, in order to keep the magnetic field divergenceless.\cite{etienne17} 
The evolution variables are linearly extrapolated to the outer boundary domain, which is causally disconnected from the interior by Adaptive Mesh Refinement (AMR) provided by the {\tt Einstein Toolkit}.\cite{ET} 
In order to prevent the force-free condition from breaking down during the evolution due to accumulation of numerical error, two supplementary conditions are imposed on $S_i$ during evolution to limit the direction and magnitude of the electric field so that it stays perpendicular to and smaller than the magnetic field: $E_iB^i=0$ and $B^2 > E^2$. 
However, this condition does not account for the current sheet (CS), which inherently forms in magnetized plasmas and it's a critical component of the magnetospheres of rapidly spinning black holes.\cite{komissarov}
We introduce a thin CS spatially localized at the equator by setting to zero the velocity perpendicular to the $z=0$ plane.
 
\emph{Initial Specifications}:
We start with the magnetospheric Wald ansatz ${A_t=0, A_i=\tfrac{B_0}{2} (g_{i\theta}+ 2 a g_{it})}$ in terms of the Kerr metric in spherical coordinates.\cite{wald} 
This vector potential creates a purely azimuthal magnetic field in Boyer-Lindquists coordinates, but in the horizon-penetrating Kerr-Schild coordinates there is an extra toroidal component $B_{\phi}$, due to the rotation of the geometry. 
The CS develops in the equatorial plane, forcing the anchored magnetic field lines to rotate with the black hole. 
There is no known exact solution for this magnetosphere, which is known as the ``ultimate Rosetta Stone" for its close connection to the BZ process.\cite{komissarov,0911.3889}

The second initial configuration is a uniform azimuthal magnetic field created by a toroidal vector potential $A_\phi=\tfrac{B_0}{2}r^2\sin^2\theta$, which in cartesian coordinates becomes $A_i=\tfrac{B_0}{2} (y, -x,0 )$
and produces a vertical magnetic field of magnitude $B_0$.
With the CS in the equatorial plane, we tilt the magnetic field off the $z$-axis with an angle $\chi$ by a rotation 
$A_i\rightarrow \tfrac{B_0}{2} (y, -x\cos \chi ,x \sin \chi)$ around the $x$-axis.  


\section{Results} 

The metric is described by the Kerr geometry in cartesian Kerr-Schild coordinates, radially shifted with $r_0=\sqrt{1-a^2}$ such that $r\rightarrow 1+r_0$, and the black hole 
inner horizon is at $r_- = 1$.
We vary the spin from $a=0.5c$ to $a=0.95c$ 
and run a suite of 15 tests, with the constant $B_0 = 0.1 M$ for the magnetic field. 
The numerical grid consists of $n=5$ levels of refinement, with resolution scaling as $\Delta x_0/2^{n-1}$ where $\Delta x_0 = 2M$, and extends to $[-50M,+50M]^2$ in the $(x,y)$ plane and to $[-100M,100M]$ in the $z$ direction.
We visualize the EM luminosity: 
$L_{EM} =\int r^2 S d \Omega$,
given by the Poynting flux across a sphere:
$S=n^i\epsilon_{ijk}\sqrt{\gamma}E^j B^k$, $n^i=\tfrac{x^i}{r}$.
In all the figures dark colors are low luminosity, bright colors are high luminosity.

\begin{figure}[th!]
\begin{center}
\includegraphics[width=1.0\textwidth]{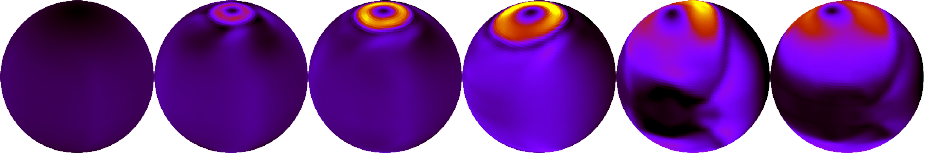}
\end{center}
\caption{Poynting luminosity of Wald magnetosphere with equatorial CS for Kerr black hole of spin $a=0.95c$ and radius $r=20M$. From left to right the time is $t=\{20, 60, 100, 140, 180, 200\}M$.}
\label{fig1}
\end{figure}

\emph{Magnetospheric Wald}: 
In Fig.~\ref{fig1} we show the time evolution of this testcase for $a=0.95c$, at $r=20M$. 
We distinguish two transient domains, between $t=\{0,50\}M$ and $t=\{150,200\}M$, when the electromagnetic field changes, and two stable phases: for $t=\{50,150\}M$ with a collimated ring-like jet structure at the poles,
and the second starting at $t=200M$ when the jet broadens and precesses.  

\begin{figure}[th!]
\begin{center}
\includegraphics[width=1.0\textwidth]{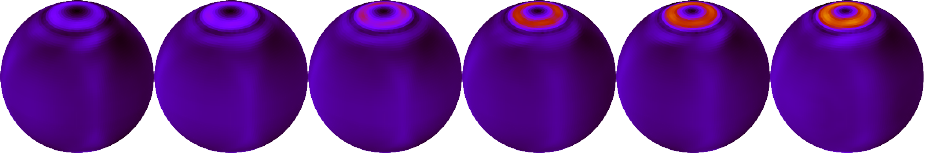}
\end{center}
\caption{Poynting luminosity of Wald magnetosphere with equatorial CS for Kerr black hole at $t=100M$ and $r=20M$.
From left to right the spins are: $a=\{0.5, 0.6, 0.7, 0.8, 0.85, 0.9\}c$.}
\label{fig2}
\end{figure}

In Fig.~\ref{fig2} we plot the Poynting flux at $r=20 M$ and $t=100M$ corresponding to the maximum amplitude and collimation of the jet. 
The intensity of the EM luminosity depends on the spin, as expected for the BZ process, however we discern the poloidal jet even at low spins, with no sign of spin threshold for jet production. 

\begin{figure}[th!]
\begin{center}
\includegraphics[width=0.67\textwidth]{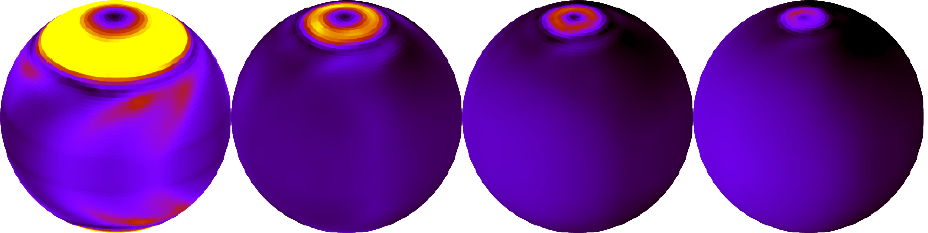}
\end{center}
\caption{Poynting luminosity of Wald magnetosphere with equatorial CS for Kerr black hole for $a=0.95c$. 
From left to right $\{r,t\}=\{10M,90M\},\{20M,100M\},\{30M, 110M\},\{40M,120M\}$.}
\label{fig3}
\end{figure}

\begin{figure}[th!]
\begin{center}
\includegraphics[width=0.84\textwidth]{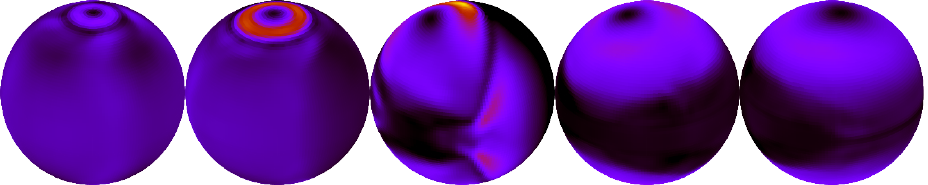}
\end{center}
\caption{Poynting luminosity of vertical magnetosphere with equatorial CS for Kerr black hole with $a=0.8c$ and $r=20M$. From left to right the time is $t=\{60, 120, 180, 240, 300\}M$.}
\label{fig4}
\end{figure}

\begin{figure}[th!]
\begin{center}
\includegraphics[width=0.67\textwidth]{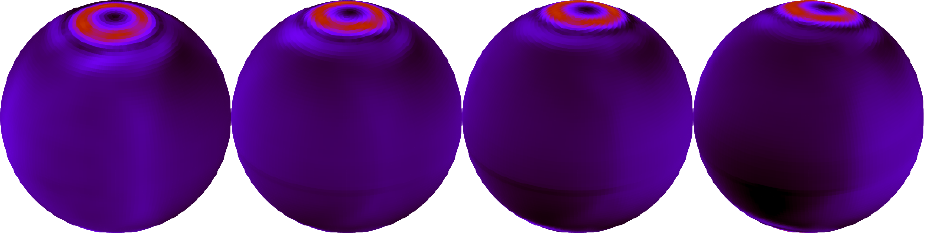}
\end{center}
\caption{Poynting luminosity of tilted magnetosphere with equatorial CS for Kerr black hole with $a=0.8c$, $r=20M$ and $t=100M$. From left to right the angle is $\chi =\{0^{\circ}, 15^{\circ}, 30^{\circ}, 45^{\circ}\}$.}
\label{fig5}
\end{figure}

\begin{figure}[th!]
\begin{center}
\includegraphics[width=0.67\textwidth]{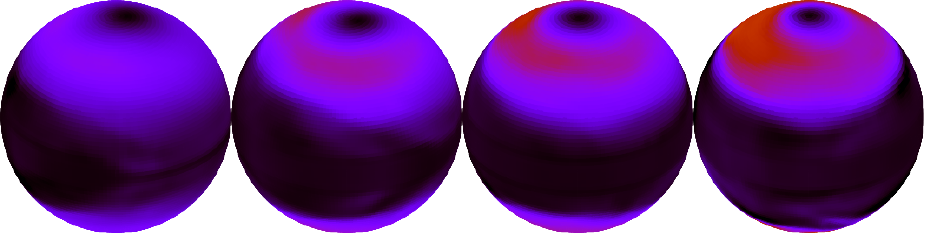}
\end{center}
\caption{Poynting luminosity of tilted magnetosphere with equatorial CS for Kerr black hole with $a=0.8c$, $r=20M$ and $t=300M$. From left to right the angle is $\chi =\{0^{\circ}, 15^{\circ}, 30^{\circ}, 45^{\circ}\}$.}
\label{fig6}
\end{figure}

In Fig.~\ref{fig3} we plot the Poynting luminosity for a Kerr hole with spin $a=0.95c$ for increasing radii $r=\{10, 20, 30, 40\}M$ at the time corresponding to the maximum amplitude and collimation of the jet. 
We see that the power of the jet scale is inversely proportional to the radius, while the collimation is directly proportional. 
\emph{Uniform Magnetosphere}: 
Fig.~\ref{fig4} shows that the time evolution of the luminosity for an initially vertical magnetosphere around a Kerr black hole is similar to the Wald magnetosphere.
We compare the luminosity for four tilt angles at $t=100M$ in Fig.~\ref{fig5} and at $t=300M$ in Fig.~\ref{fig6}, and we see that it increases with the tilt angle. 

\section{Discussion}
We find that the geometry brings the black hole magnetosphere to a stability domain with a poloidal, precessing jet.
We see two stable phases: early on, the jet forms a collimated, ring-like structure at the poles, while at late times, the jet broadens and precesses.
Both the spin and the tilt angle affect the amplitude but not the shape of the jet, with no indication of high spin threshold for jet formation. 
Visualizations are available at {\tt https://github.com/mbabiuc/KerrMagnetosphere}.

\section*{Acknowledgments}
This work supported by the National Science Foundation Award No. IIA-1458952.
We gratefully acknowledge Z. Etienne and M. B. Wan for valuable discussions.
The computer simulations were carried on the BigGreen Cluster at Marshall University.

\end{document}